\documentclass{article}

\usepackage{arxiv}

\usepackage[utf8]{inputenc} 
\usepackage[T1]{fontenc}    
\usepackage{hyperref}       
\usepackage{url}            
\usepackage{booktabs}       
\usepackage{amsfonts}       
\usepackage{nicefrac}       
\usepackage{microtype}      
\usepackage{lipsum}		
\usepackage{graphicx}

\title{Eight years of homicide evolution in Monterrey, Mexico: a network approach}

\author{{Rodrigo Dorantes-Gilardi} \\
	Centro de Estudios Internacionales \\
	El Colegio de M\'exico\\
	Mexico City\\
	\texttt{rdorantes@colmex.mx} \\
	\And
	{Diana Garc\'ia Cort\'es} \\
	Computational Genomics Department\\
	National Institute of Genomic Medicine\\
	Mexico City \\
	\texttt{dgarcia@inmegen.edu.mx} \\	
	\And
	{Hiram Hern\'andez Ramos} \\
	Facultad de Ciencias\\
	Universidad Nacional Aut\'onoma de M\'exico\\
	Mexico City\\
	\texttt{hiram.hernandez@ciencias.unam.mx} \\	
	\And
	{Jes\'us Espinal-Enriquez} \\
	Computational Genomics Department\\
	National Institute of Genomic Medicine\\
	Mexico City \\
	\texttt{jespinal@inmegen.gob.mx} \\}

\renewcommand{\shorttitle}{\textit{arXiv} Template}

\hypersetup{
pdftitle={A template for the arxiv style},
pdfsubject={q-bio.NC, q-bio.QM},
pdfauthor={David S.~Hippocampus, Elias D.~Striatum},
pdfkeywords={crime networks, homicide dynamics, spatial networks, Mexico's drug war}}

\begin{document}
\maketitle

\begin{abstract}
	Homicide is without doubt one of Mexico's most important security problems, with data showing that this dismal kind of violence sky-rocketed shortly after the war on drugs was declared in 2007. Since then, violent war-like zones have appeared and disappeared throughout Mexico, causing unfathomable human, social and economic losses. One of the most emblematic of these zones is the city of Monterrey, a central scenario in the narco-war. Being an important metropolitan area in Mexico and a business hub, Monterrey has counted hundreds to thousands of casualties. In spite of several approaches being developed to understand and analyze crime in general, and homicide in particular, the lack of  accurate spatio-temporal homicide data results in incomplete descriptions. To better understand the underlying mechanisms by which violence has evolved and spread through the city, here we propose a network-based approach. For this purpose, we define a homicide network where nodes are geographical entities that are connected through spatial proximity and crime similarity. Data is taken from a crime database spanning 86 months in the Monterrey metropolitan area, containing manually curated geo-located and dated homicides, as well as from Open Street Map for urban environment. Under this approach, we first identify independent crime sectors corresponding to different connected components. Each of these clusters of crime presents crime evolution similar to the one at state and national levels. We then show how crime spread from neighborhood to adjacent neighborhoods when violence was mainly cartel-related and how it was chiefly static at a different time. Finally, we show a relation between homicidal crime and urban landscape by studying the distance of safe and violent neighborhoods to the closest highway and by studying the evolution of highway and crime distance over the cartel-related years and the following period. With this approach, we are able to describe more accurately the evolution of homicidal crime in a metropolitan area.
\end{abstract}

\section*{Introduction}
\subsection*{Violence in Mexico during the drug war}
In an attempt to legitimate his questionable victory on the 2006 presidential elections, former Mexican president, Felipe Calderon Hinojosa (FCH), launched an intervention of the Mexican Army and Federal Police into the State of Michoac\'an, to face  the drug cartels that operated in that region~\cite{rose2007}. ``Operaci\'on Michoac\'an'' was the starting point of FCH's Drug War~\cite{aranda2013}. This battle  between Mexican security forces and drug cartels spread the violence throughout the whole country and homicide rates scaled to levels never seen before in Mexico~\cite{espinal2015}.

The dramatic increase of drug-related homicides during FCH's drug war, brought a wave of violence as a collateral effect to the aforementioned war. During the administration of the next president, Enrique Pe\~na Nieto (EPN), there was an apparent period of lower homicide rates. That period corresponded to his first three years of government. However, at the beginning of 2015 a renewed wave of violence emerged at a national level and has not decreased since~\cite{rosen2016}.

FCH's drug war also generated battles between drug cartels to control broad territories. In the year of 2010, there was a schism between the Gulf Cartel, one of the most important criminal groups in Mexico, and the ``Zetas'' band, the former armed force of the Gulf Cartel. This struggle took place in the northern states of Tamaulipas and Nuevo Le\'on. The violence triggered by this separation caused the highest homicide rates in the history of several cities in the region such as Monterrey~\cite{espinal2015}, and, more generally, in the Monterrey Metropolitan Area (MMA), the second urban area in Mexico, and an important industrial hub.

The rise of violence during the second part of EPN's administration, was not only due to the drug war, it has been demonstrated that drug-band related homicides have a strong influence in general homicides~\cite{cohen1998}. The increasing homicides rates have thus permeated in Nuevo Le\'on and the MMA. 

\section*{Geo-social MMA Background}

	\begin{figure*}[h!]
  	\centering
  	\includegraphics[width=.75\textwidth]{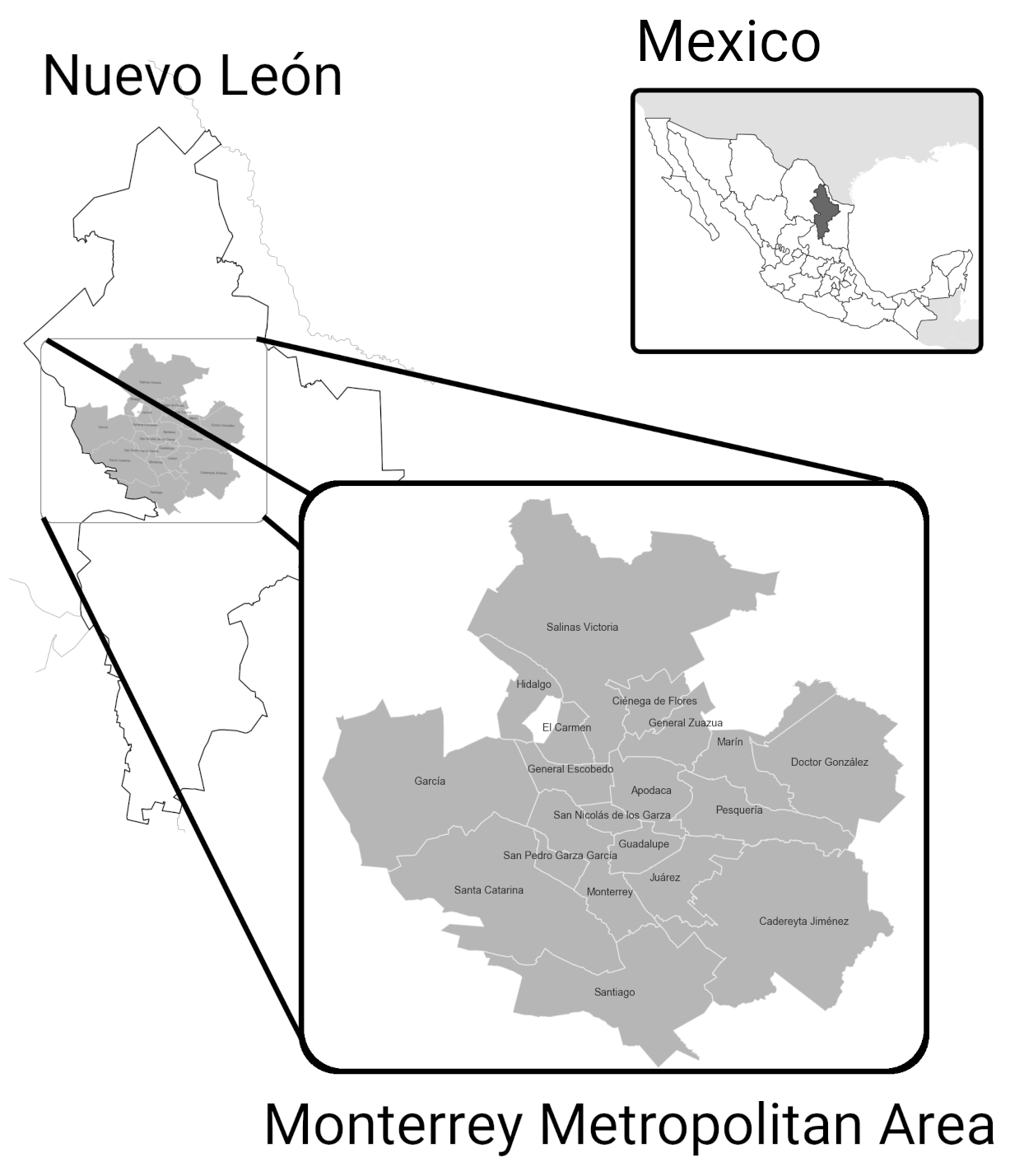}  
	\caption{\textbf{Political division of MMA.} This map contains the names of municipalities in the MMA and their location inside the state of Nuevo León. The location of Nuevo Le\'on within Mexico is also displayed. }
	\label{names}
	\end{figure*}

MMA is composed by 19 municipalities, namely, Monterrey, Guadalupe, San Nicol\'as de los Garza, Apodaca, San Pedro Garza Garc\'ia, Santa Catarina, General Escobedo, Ju\'arez, Hidalgo, Santiago, Cadereyta Jim\'enez, García, Salinas Victoria, Pesquer\'ia, Ci\'enega de Flores, General Zuazua, Mar\'in, Carmen, y Doctor Gonz\'alez. The number of neighborhoods within these municipalities comes to $2,691$, where more than 4.5 Million people live (Figure~\ref{names}).

Neighborhoods can tile up a city in the way that each tile is a village of its own. Studies about cities in the United States have shown that crime is highly concentrated in some neighborhoods within cities~\cite{rosenfeld2007impact}. However, a typical urban tiling comprises hotspot neighborhoods bordering low crime-rate ones~\cite{zorbaugh1983gold}. It is not uncommon in Latin America to have extreme cases of slums next to luxurious neighborhoods separated typically by urban infrastructure. MMA is not the exception; while San Pedro Garza Garc\'ia is the second municipality in terms of human development index (HDI=0.87 in 2015~\cite{snim}), the municipality of Santa Catarina (Northern border of San Pedro Garza Garc\'ia) has neighborhoods with important levels of poverty. Moreover, neighborhoods have been shown to be interdependent in terms of what happens in one affects the other~\cite{tita2009crime}. In particular, crime shows diffusion processes across neighborhoods~\cite{sampson2012great, zeoli2014homicide}.

\section*{Approaches to understand violence}

There have been many attempts to describe, understand, predict or control the dynamics and spread of conflict and gang-related violence: from literature-based approaches \cite{espinal2015}, data-mining-based network inference \cite{brantingham2011, kitts2014}, reaction-diffusion equations \cite{zeoli2014homicide}, to combined methods \cite{papachristos2013, papachristos2009}.  Understanding temporal and spatial evolution of homicides in a metropolitan area is of utmost importance to alleviate and diminish said violence. 

A crucial step in the analysis and development of accurate models of homicide dynamics is data collection, in particular, availability of geolocated data and precise dating of  events is required. In this sense, works such as the one developed by Oliveira \cite{oliveira2018} have collected crime data of robbery and burglaries. Other approaches have used homicide data, however, the level of granularity for these data is in the best cases by municipality \cite{espinal2015, pina2019exploring}, or by country \cite{richardson1941}. In both cases: lack of precision in geolocated data and large time-steps limit the description of a complex phenomenon such as the spread of crime-related homicides throughout a metropolitan area. 

Newspaper \textit{El Norte} has documented the homicide violence in the MMA from the year of 2011 until February of 2018. Its database reports daily homicides with number of casualties for each event and the associated longitude and latitude coordinates. To our knowledge, this is the most accurate and comprehensive geo-located homicide database for any place in Mexico.

\section*{Outline}

In this work, we used geo-located homicide data to analyze the structure and dynamics of homicide-related violence in the MMA. Data was taken from \textit{El Norte} newspaper daily-updated database (ENDB). We developed a spatial and temporal analysis of the homicides in the MMA.

We first analyzed the time series of homicides at municipality (city), locality, and neighborhood level. By means of neighborhood-derived analysis, we assembled a network where neighborhoods are connected if they are adjacent and if there was at least one homicide during a given time window. Said networks were intersected by periods.  

Second, to detect whether or not violence was correlated in a temporal fashion, neighborhood networks were build if two places had at least one event during the same week, or if events were separated by one week. Finally, we correlated the places in which homicides took place and Open Street Maps (OSM) locations. We observed the most frequent places close to homicides, to assess whether or not a tendency appeared.

The approaches used here allow us to answer different questions related to the dynamics and structure of homicidal violence: Is violence in MMA spreading through time? Is violence spreading through space? Do municipalities in the MMA present a concerted pattern of violence or is there a correlation among them? Is violence bursting simultaneously or a time-window appears between violence in different places? Are there features of the urban environment related to the location of homicides?. 

The analysis of close neighborhoods showed that during 6-month periods there is only links between Monterrey Downtown Neighborhood and its closest neighbors. The within-week networks and 1-week-separated correlation networks showed that the most correlated places are geographically apart, contrary to the already known fact that physical proximity determines the spread of violence. Finally, the OSM data shows that homicides correlate with locations close to highways and freeways, but more importantly, main roads in MMA segregate homicidal neighborhoods from the non-violent ones. Additionally, the political division of some municipalities, and the roads appear to constrain the areas in which homicides occur.

With this approach, we conclude that the dynamics of homicides in MMA is not similar to other places in which data exist. We argue that the evolution of those dynamics were strongly influenced by the FCH's drug war. Here we have an accurate description of the structure and dynamics of homicide crime in a large metropolitan area.

\section*{Results}

\begin{figure*}[h!]
  	\centering
  	\includegraphics[width=.98\textwidth]{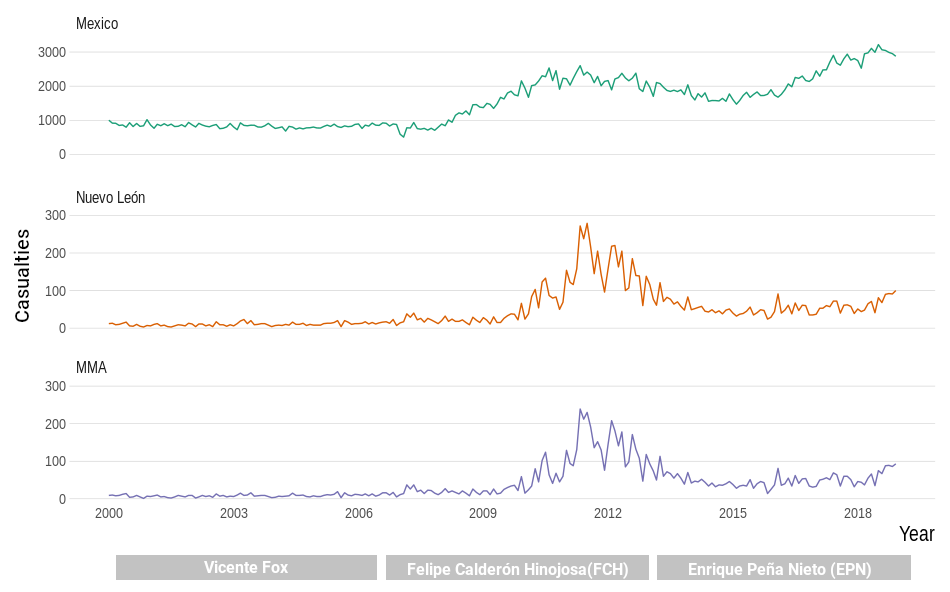}  
	\caption{\textbf{Timeseries of homicides in Mexico, Nuevo Le\'on State, and the Monterrey Metropolitan Area (MMA).} In this plot, the monthly number of homicides for those three levels of granularity are described. It can be observed at the beginning of the FCH's presidency, the substantial increase in homicide rate in the three levels. The second increase in homicides in Mexico did not follow the same trend in Nuevo Le\'on nor MMA, however, in both places violence increased. }
	\label{mapa}
	\end{figure*}

Figure~\ref{mapa} shows the monthly official data of homicides in Mexico, the Mexican State of Nuevo León, and the area given by the Monterrey Metropolitan Area (MMA) from 1990 until 2019. As it can be observed there is a steady increase in the number of homicides starting in 2007 until 2012 at national level, and a second one from 2016 to 2018. The year of the first increase in homicides (2007) coincides with the start of the war on drugs started by president Felipe Calderón. In the case of Nuevo León, the first steady increase in homicides happens almost three years later, at the end of 2009 and reaches almost 10\% of all homicides in Mexico in 2011, the peak of the narco war in Monterrey. The permeation of the drug war reached MMA at the end of 2009 and directed the fluctuation of homicides at a State level, as it can be seen in Figure~\ref{mapa}. We also observe a second smaller increase in Nuevo León after 2014, posterior to a steady decrease in homicides during the period 2012--2014.

		\begin{figure*}[h!]
  	\centering
  	\includegraphics[height=18cm]{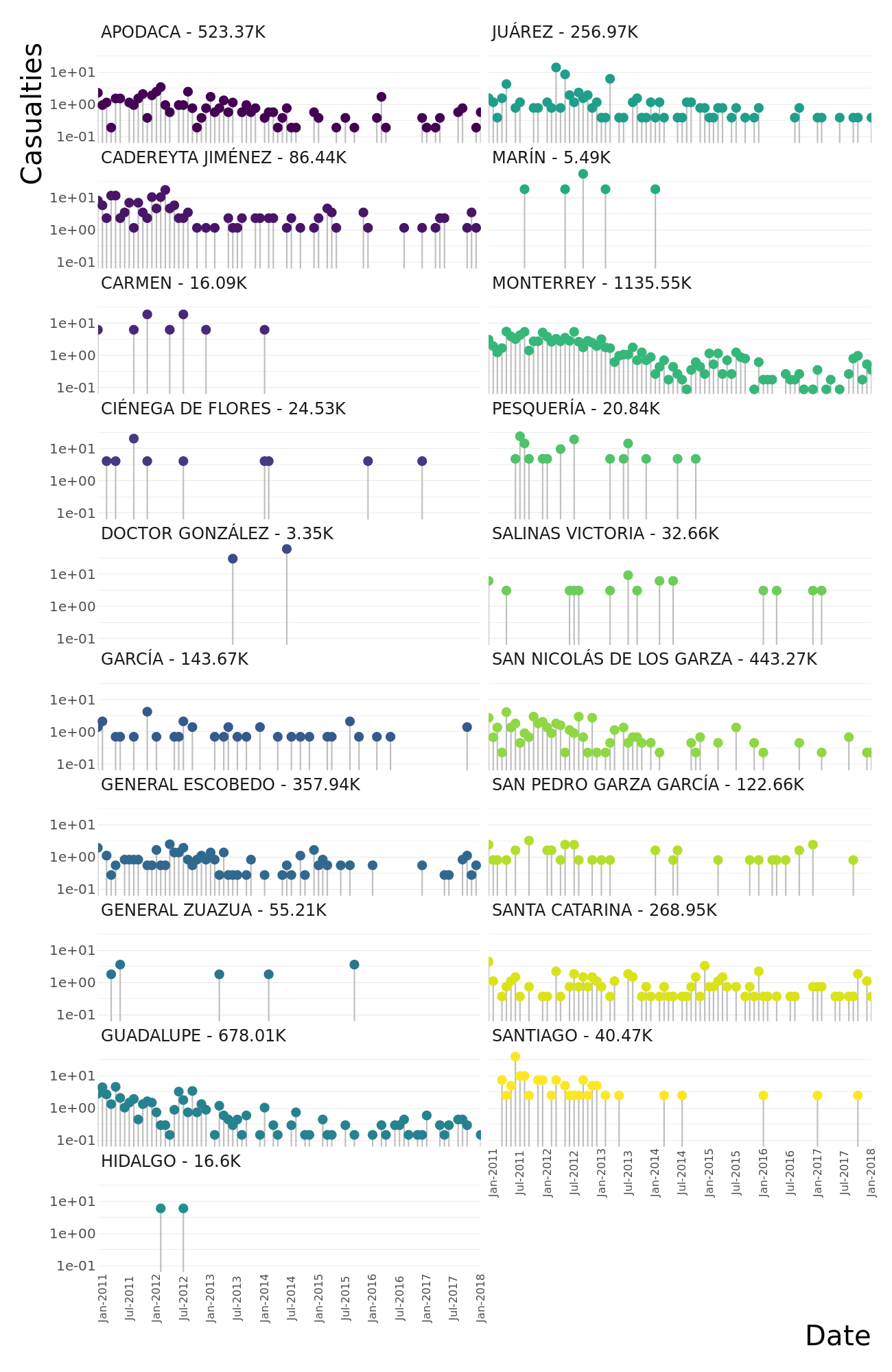}  
	\caption{\textbf{Timeseries of homicides  by week in the cities of MMA}. This \textit{lollipop plot} shows the number of casualties for each municipality of MMA during the period under study as well as the population of each municipality (next to name). The upper right panel shows the cumulative number of homicides in the ``El Norte Data Base''. As it can be observed, the trend of ENDB data coincides with MMA homicides reported by INEGI in Figure \ref{mapa}.}
	\label{paleta}
	\end{figure*}

Figure~\ref{paleta} shows the casualties by municipality of the MMA in the form of a lollipop plot. Each point represents an aggregation of the monthly casualties per one-thousand inhabitants in each municipality. There is no homogeneity in homicide rates across municipalities, revealing crime focal points located within some municipalities during this period. This is mainly the case of Apodaca, Monterrey, Cadereyta Jiménez, San Nicolás de los Garza and Guadalupe. The distribution of homicide events is denser around the 2011--2012 period, decreasing around 2014 and increasing again after 2016, coinciding with the homicide trend at a State level from Figure~\ref{mapa}. 

Of particular interest is the case of Santa Catarina municipality: Despite it not being one of the top places in terms of number of casualties in MMA, it consistently presents homicides during the whole period, even during the last months of 2017 and beginning of 2018.

	\begin{figure*}[h!]
  	\centering
  	\includegraphics[width=.9\textwidth]{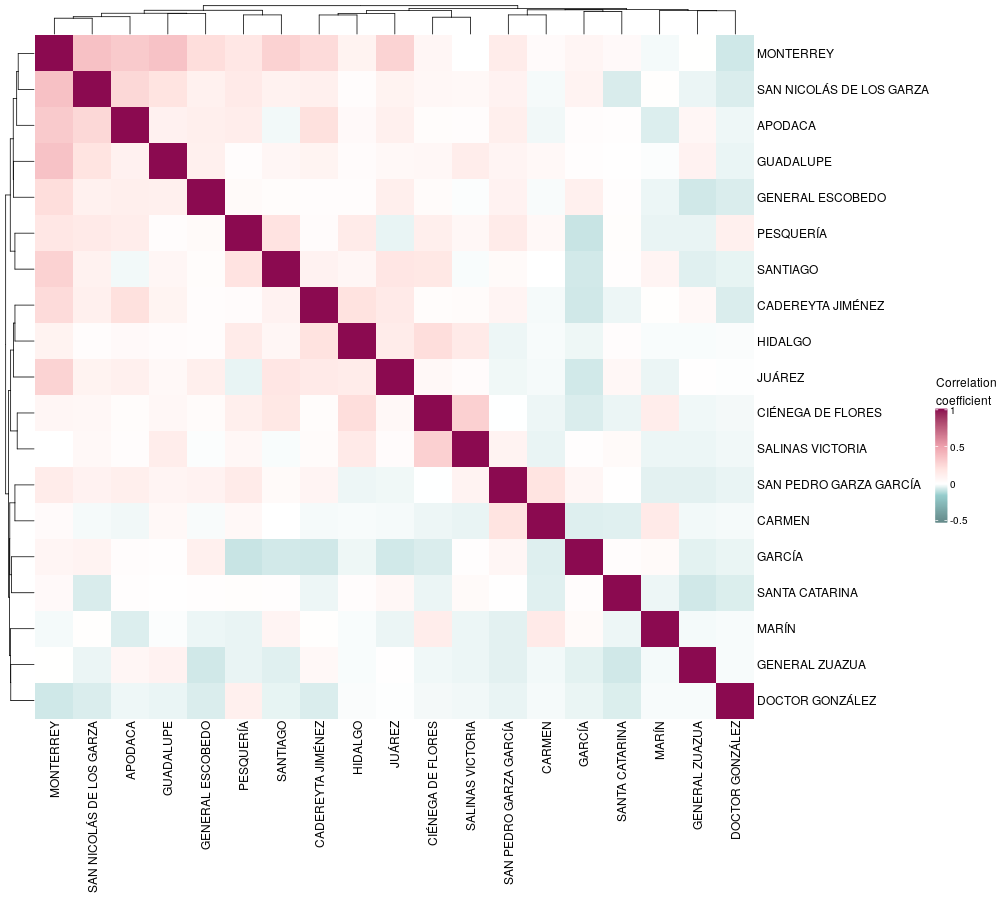}  
	\caption{\textbf{Correlation of homicides in the cities of MMA}. This heatmap represents the Pearson correlation coefficient of the homicide time series of the 19 municipalities that compose the MMA. Pink squares show positive correlation, meanwhile negative correlations are depicted in blue.}
	\label{heatmap}
	\end{figure*}

The homicide time-series were used to correlate violence across municipalities across the 86-month period on a weekly basis (Figure~\ref{heatmap}). The highest correlated city-pair is Monterrey and San Nicolás de los Garza, meanwhile García and Pesquería are the highest anti-correlated municipalities. We assigned a significance score via Z-scores to the correlation values by means of a null-model built from reshuffled iterations of the casualties by municipality. Significantly correlated (Z-score $> 3$) municipality pairs are displayed in Table~\ref{table:correlation}.

\begin{table}[h!]
\caption{Pearson correlation coefficient and Z-score of weekly aggregated casualties at a municipality level}
\centering
	\begin{tabular}{llrr}  
        \toprule
          Municipality 1 &  Municipality 2 &  PCC & Z-score\\ \midrule
       MONTERREY	&	SAN NICOLÁS DE LOS GARZA	&	0.3581	&	6.4388	\\
GUADALUPE	&	MONTERREY	&	0.3557	&	6.3556	\\
APODACA	&	MONTERREY	&	0.317	&	5.6961	\\
CIÉNEGA DE FLORES	&	SALINAS VICTORIA	&	0.2893	&	5.4055	\\
MONTERREY	&	SANTIAGO	&	0.2839	&	5.0035	\\
JUÁREZ	&	MONTERREY	&	0.2788	&	4.8732	\\
APODACA	&	SAN NICOLÁS DE LOS GARZA	&	0.2547	&	4.6693	\\
CADEREYTA JIMÉNEZ	&	MONTERREY	&	0.2422	&	4.2779	\\
CIÉNEGA DE FLORES	&	HIDALGO	&	0.2263	&	4.143	\\
GENERAL ESCOBEDO	&	MONTERREY	&	0.2291	&	4.0746	\\
CADEREYTA JIMÉNEZ	&	HIDALGO	&	0.2053	&	3.8278	\\
APODACA	&	CADEREYTA JIMÉNEZ	&	0.2149	&	3.8185	\\
PESQUERÍA	&	SANTIAGO	&	0.2027	&	3.7373	\\
GUADALUPE	&	SAN NICOLÁS DE LOS GARZA	&	0.2014	&	3.6547	\\
CARMEN	&	SAN PEDRO GARZA GARCÍA	&	0.1992	&	3.4842	\\
JUÁREZ	&	SANTIAGO	&	0.1829	&	3.1747	\\
MONTERREY	&	PESQUERÍA	&	0.1756	&	3.0951	\\
CIÉNEGA DE FLORES	&	SANTIAGO	&	0.1714	&	3.0139	\\
\bottomrule
	\end{tabular}
	\label{table:correlation}
\end{table}

\subsection*{Spatial distribution of homicides}

	\begin{figure*}[h!]
  	\centering
  	\includegraphics[width=.7\textwidth]{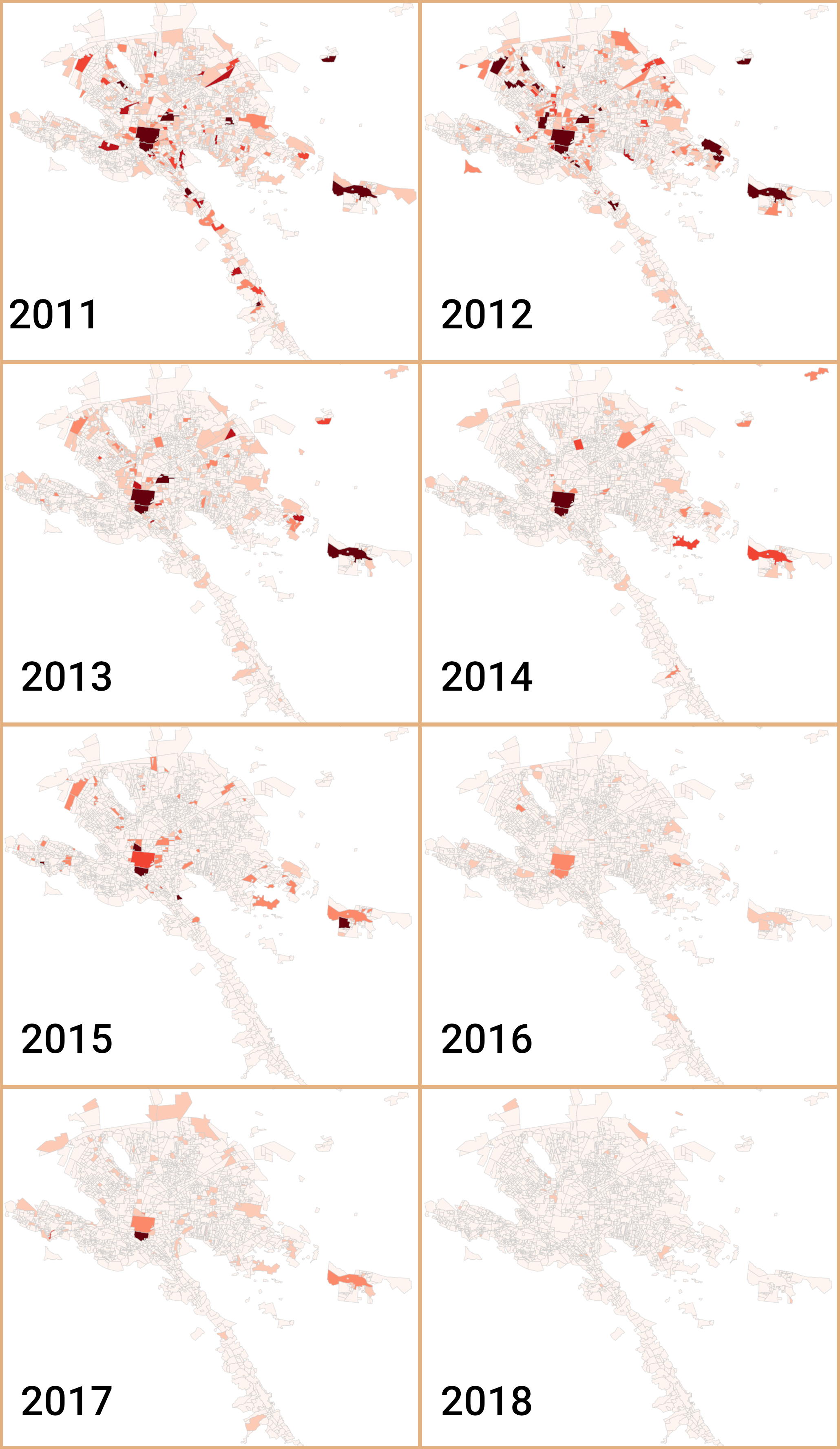}  
	\caption{\textbf{Geospatial yearly distribution of homicides in MMA}. In these maps, neighborhoods are depicted according to the number of homicides that took place there. light colors represent lower number of casualties, meanwhile dark colors take account for higher homicide numbers.}
	\label{mapaheat}
	\end{figure*}	
	
In order to have a more accurate description of the dynamics of homicidal violence, we observed the neighborhoods in which events occurred. In Figure~\ref{mapaheat}, we depicted the number of casualties per neighborhood for  all years under study. As it can be appreciated, the patterns of violence change over the period of time, and space. Importantly, we can observe the above-mentioned decrease in homicides after 2016.

\subsection*{Spatial evolution of homicides}
In order to better understand the change in homicide patterns across time, we use a finer level of granularity of MMA: neighborhoods. We counted a total of $2,699$ neighborhoods in $19$ municipalities, where $769$ of these neighborhoods have at least one homicide in the January 2011--February 2018 time-window. In other words 28\% of all neighborhoods in MMA suffered a homicide within their boundaries. This contrasts with the known property of focalized crime on a small number of neighborhoods within a city~\cite{rosenfeld2007impact}.

To look at the evolution of homicides across space, we constructed a homicide network representing the spatial configuration of violence within a given year, for each year of our time-window. In the homicide network nodes are neighborhoods and two nodes are connected by an edge if (i) they are spatially adjacent neighborhoods (their respective polygons are adjacent), and (ii) there was at least one homicide in the two neighborhoods.

		\begin{figure*}[h!]
  	\centering
  	\includegraphics[width=.7\textwidth]{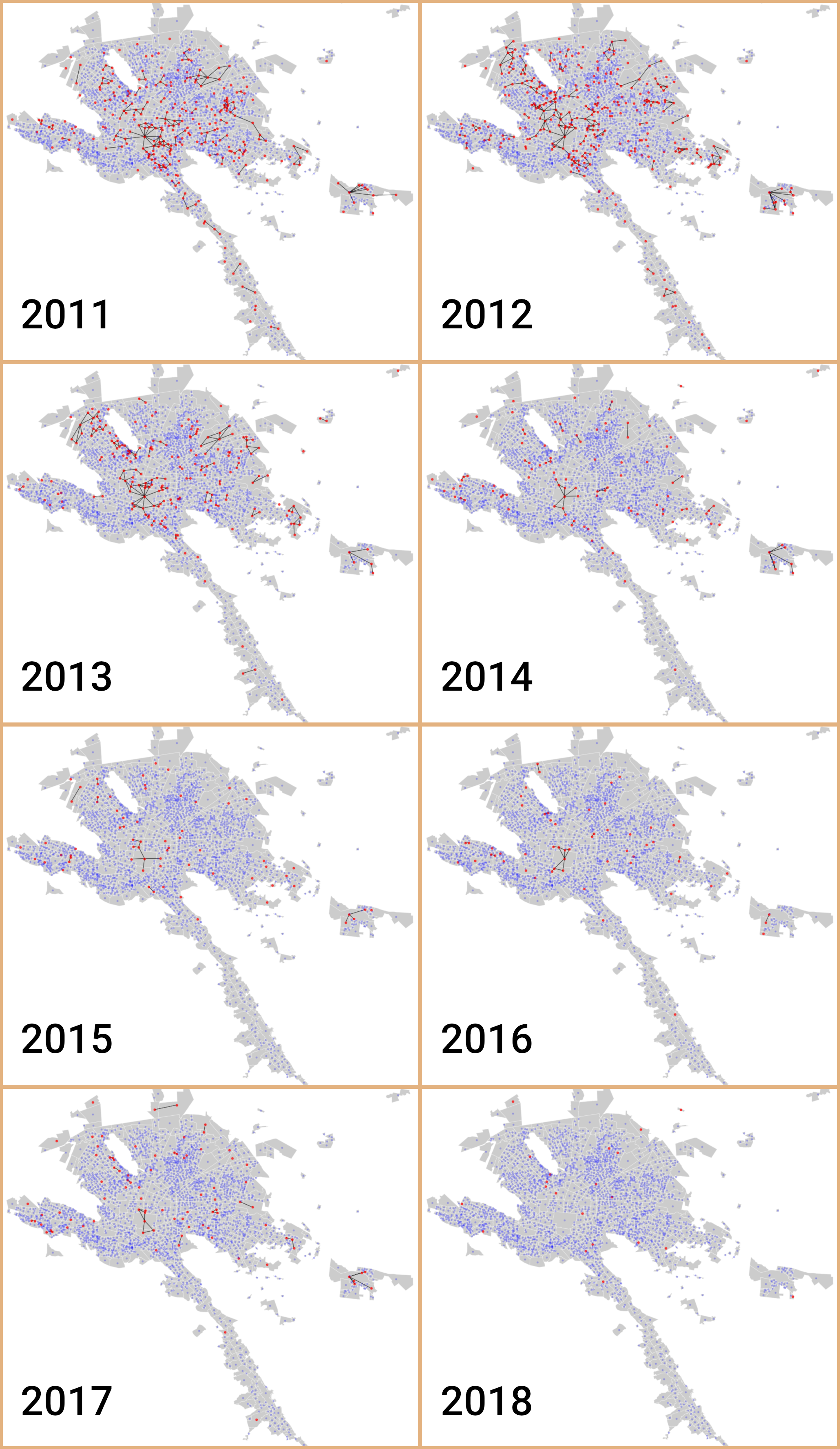}  
	\caption{\textbf{Geospatial distribution of homicides in MMA during the eigth years}. This network representation shows the clusters formed by all neigborhoods that have at least one homicide and are geographycally together. Each color represents a network component. The bottom left table shows the municipalities of the MMA that belong to each cluster, and the number of neighborhoods belonging to that city.}
	\label{yearly-crime-networks}
	\end{figure*}

The analysis of each yearly crime network shows the spatial dynamics of adjacent neighborhoods in terms of homicide (Figure~\ref{yearly-crime-networks}). The global spread of violence in distant focal points is seen across all years (scattered red points, neighborhoods with at least one homicide that year), suggesting a division in crime sectors (blue points are neighborhoods without homicides). The structure of crime sectors however, varies greatly between years. Years 2011 and 2012 have crime networks with many connected components (45 and 50, respectively) suggesting a spread of homicides between a usually violent neighborhood to its adjacent neighborhoods. The number of connected components decreased steadily until 2014 onwards (18), where a pattern of isolated violent neighborhoods emerges. Considering that the war on drugs climaxed in the MMA in 2011 and decreased in late 2012, we can expect that the spreading of crime during these years is due to the unique properties of a war-like state. After 2014 the number of connected components stabilized until 2017, where we see a 2-fold increase.

				\begin{figure*}[h!]
  	\centering
  	\includegraphics[width=0.75\textwidth]{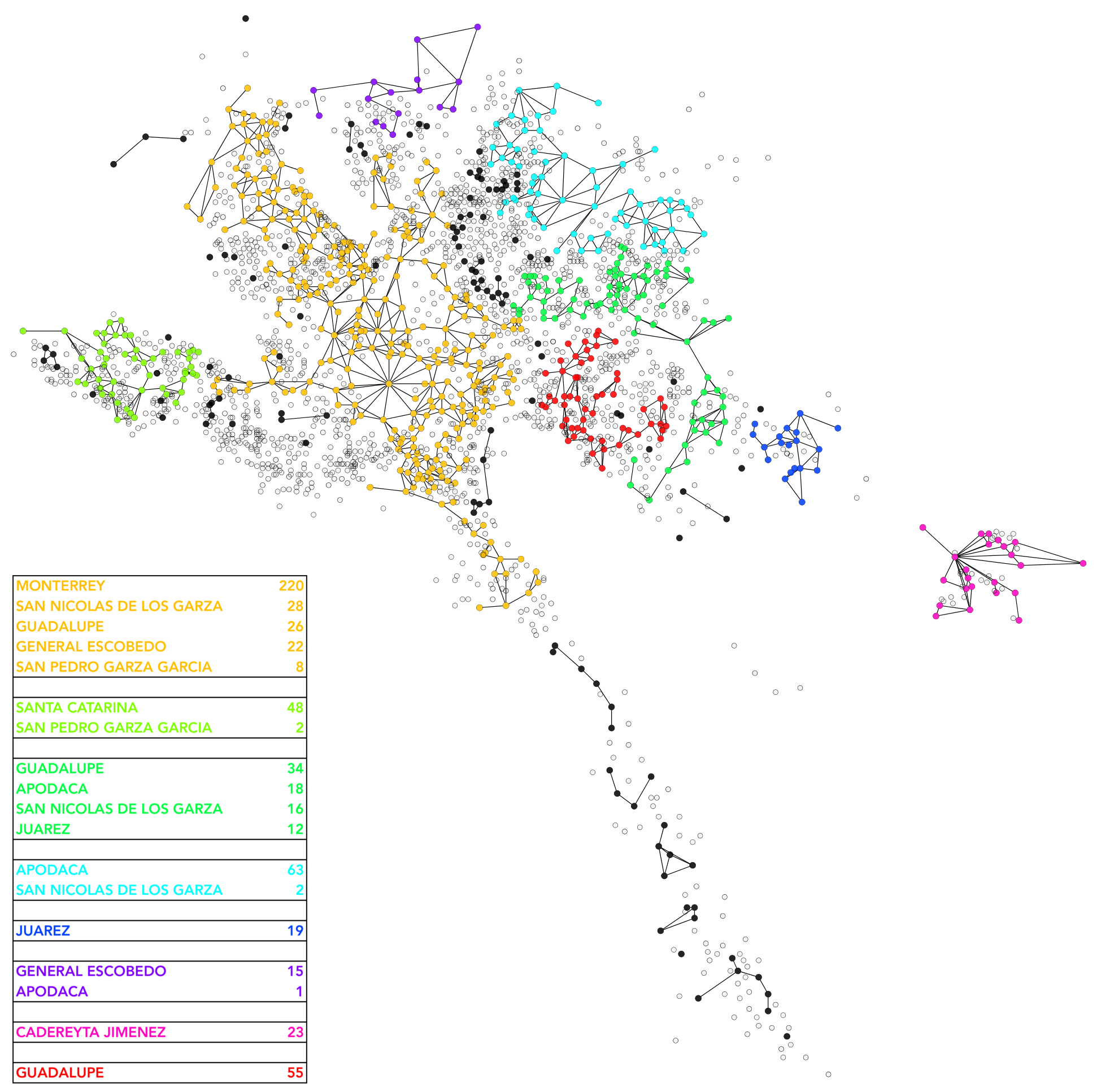}  
	\caption{\textbf{Network composed by adjacent violent neighborhoods during the entire period under study}. Each color represents a  connected component with more than 10 edges. The bottom left table indicates the municipalities that belong to each component. Black dots are neighborhoods with at least one homicide but not connected to large components.}
	\label{crime-network}
	\end{figure*}

In order to have a global view of the homicidal crime, we constructed a network using the whole-period window. There, nodes are connected if they are adjacent and have at least one homicide during the 86 months. Figure~\ref{crime-network} represents such a network, where nodes are colored by connected component with more than 10 edges (main connected components), white nodes are neighborhoods without crime (not in crime network), and black nodes belong to connected components with less than 10 edges.

Looking at the composition of the nodes, each one of the eight main connected components adheres to a main municipality strongly. For instance, yellow nodes belong to a main component with principal municipality Monterrey, containing 72\% of the nodes. Red connected component has all its nodes in the municipality of Guadalupe, to the east. Green component in the west has 55 of its nodes in Santa Catarina.

It is worth noticing cases such as the border between Santa Catarina and San Pedro Garza García municipalities (Green and white nodes in the west). Despite the fact that several neighborhoods of both cities share a border, none of these share an edge. Interestingly, safe neighborhoods seem to be located around the municipality boundaries thus disconnecting the main connected components.

	
\subsection*{Homicides are related to urban environment}
As we saw in the previous result, crime connected components or crime sectors are usually located within municipalities. Although interesting by itself, the fact that municipality borders can act as a separator between crime sectors leads to ask whether there are other barriers to crime in the MMA. 

The relevance of high-speed roads or highways in crime has been previously shown to be an important factor~\cite{kim2018physical}. Specifically in the context of a time window comprising part of a drug war, highways are of high relevance: they are usually the spots where bodies were abandoned or displayed, persecutions between criminals and police/military take place on highways, etc. Moreover, highways crossing a city act as an urban boundary. They separate neighborhoods, and municipalities, and can separate social and urban landscapes.

In order to include highways in our data, specifically in order to observe the proximity of crime events near a highway and also analyze the distance of violent neighborhoods from them, we obtained all Open Street Map (OSM) data points related to highways in the MMA.

		\begin{figure*}[h!]
  	\centering
  	\includegraphics[width=.5\textwidth]{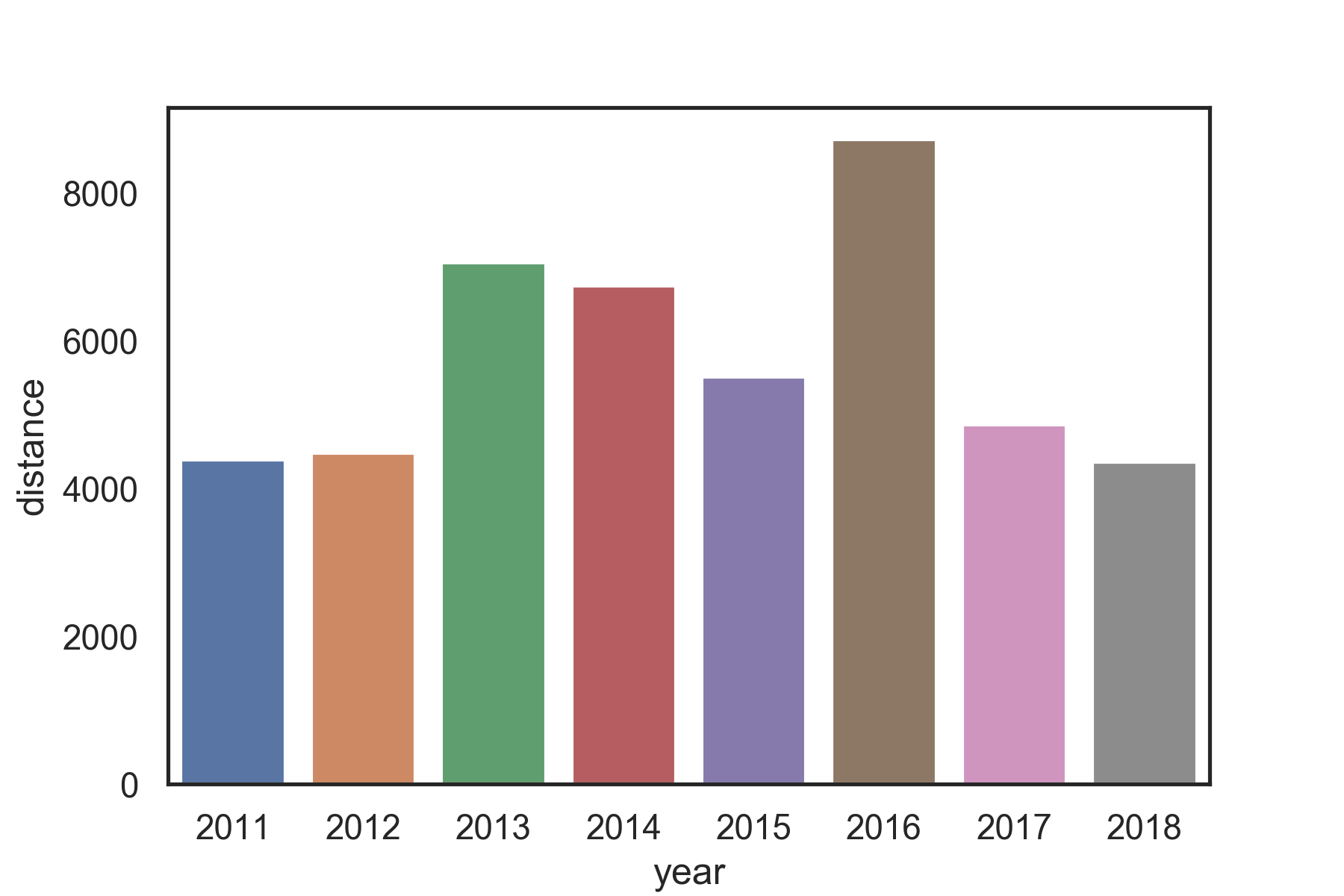}  
	\caption{\textbf{99th percentile in meters of distance from highway to crime by year}. }
	\label{distance-crime}
	\end{figure*}

As we see in Figure~\ref{distance-crime}, distance from crime to highway varies between years. In order to ignore outlier points we consider the 99th percentile (distance such that 99\% of crimes are closer than or equal to a highway) over the eight years. The two years related to the climax of the war on drugs (2011--2012) show the shortest distances to highways (4379 and 4475 meters, respectively). The time period between 2013--2016 shows longer distances to highways, and a decrease in distance is observed again in 2017.

\begin{figure*}[h!]
  	\centering
  	\includegraphics[width=0.75\textwidth]{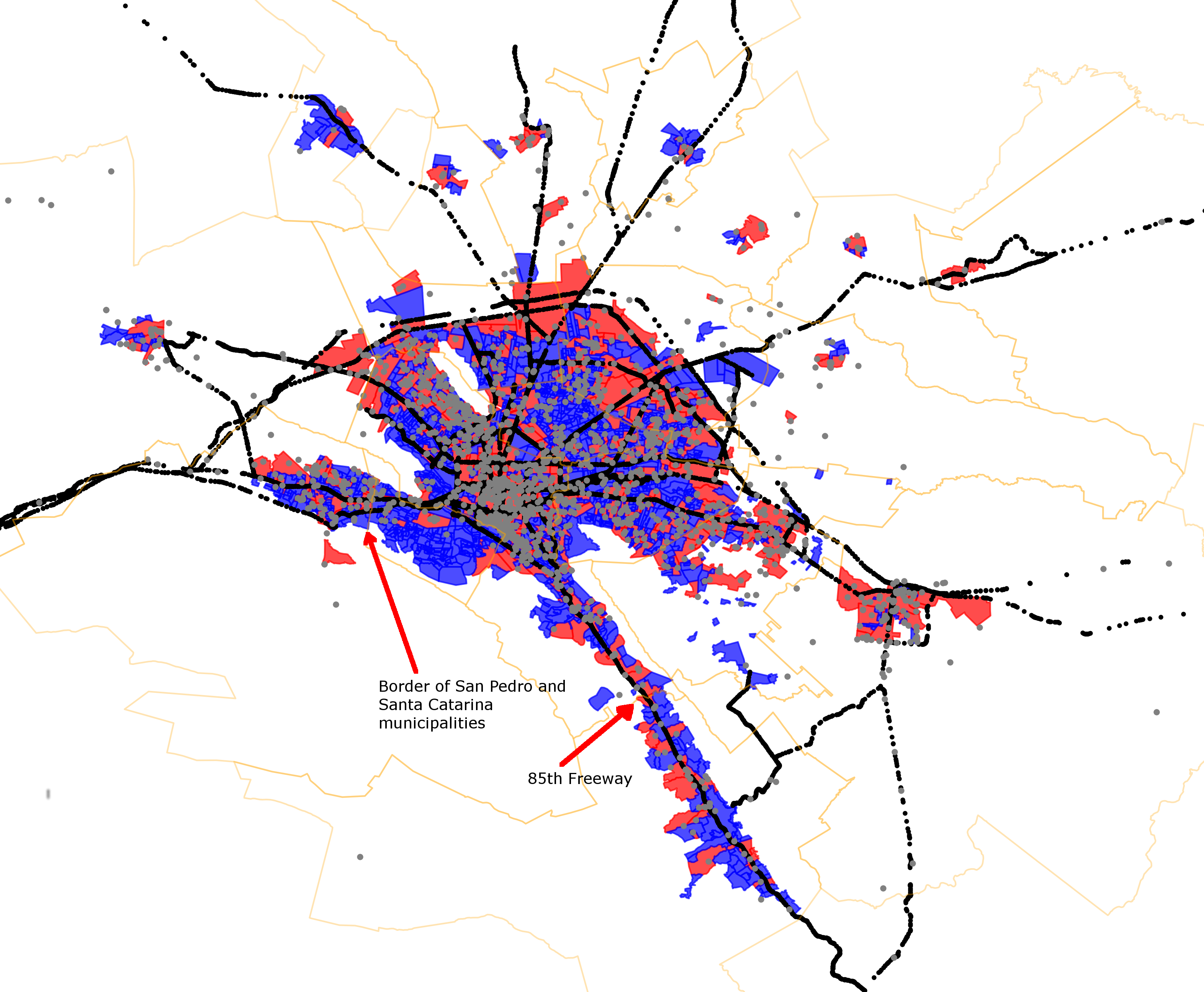}  
	\caption{\textbf{Geospatial distribution of homicides in MMA with highways}. }
	\label{roads}
	\end{figure*}

In general, highways appear to be part of a backbone of violence as many of the homicides lay directly on the highway paths (Figure~\ref{roads}). It can also be appreciated how the political division together with the highways could be forming the patterns between violent and non-violent neighborhoods (red and blue neighborhoods, respectively). Indeed, the polygons resulting from the intersection between municipalities and highways appears to unveil crime patterns and should be further investigated. One of the most evident effects of this separation is again San Pedro Garza García and Santa Catarina municipalities.

	\section*{Discussion}
\subsection*{Spatiotemporal dynamics of homicide}
Regarding the temporal evolution of crime, an interesting result came from data of Santa Catarina municipality. As mentioned in the results section, Santa Catarina is not the most violent municipality in MMA, but this place has a monotonous homicide rate, even during the years of low homicide rates in MMA. This behavior contrasts with the rest of municipalities of the MMA: It can indeed be appreciated in Figure~\ref{heatmap}, where Santa Catarina's correlations are the lowest for the entire set.

On the other hand, the other municipalities with high homicide rates have a similar behavior, with an important increase in 2011--2012, short after decreasing until 2017, where violence raised again, but at lower levels.

\subsection*{Socio-economic segregation of homicide violence}
Opposite to Santa Catarina, San Pedro Garza García municipality has a very low homicide rate (150 and 44 homicides, respectively). These two cities are separated by one avenue: Cromo Street. In the side of San Pedro there were no homicides, as opposite to Santa Catarina, where several neighborhoods suffered from homicidal violence. This contrast can be attributed to the difference in the GDP between both places. San Pedro Garza García is the second city with the highest GDP in Mexico~\cite{inegi}, meanwhile some neighborhoods in Santa Catarina have important development lag. According to the last economical survey (2015), the GDP  \textit{per capita} in San Pedro Garza García was \$25,636 USD, meanwhile that of Santa Catarina was \$10,783 USD.

Similar to this behavior related to revenue per capita,  other municipalities such as San Nicolás de los Garza, have a small number of neighborhoods with homicides compared to the rest of municipalities within the city. However, in the case of San Nicolás de los Garza, homicides occurred chiefly in its periphery. This can be observed from Figure~\ref{roads}, several neighborhoods on its central area are in blue, meanwhile the borders are in red (north-east of MMA).

Concomitant with the political borders of municipalities, highways also appear to depict lines of segregation between secure and violent neighborhoods. The remarkable difference between neighborhoods at opposite sides of a highway may reflect the presence of security forces in only one of both sides. Further investigation regarding this topic is needed to reach a solid conclusion, however, we discuss here the empirical observation.

\subsection*{Networks and geopolitical division of homicide}
The crime network constructed from joining adjacent neighborhoods if they had at least one homicide during the time window, clearly shows that homicide events are linked to the municipality in which the homicide occurred. Similar to the aforementioned relationship between GDP and violence, here we observe that homicide regions are strongly segregated by the political division of the MMA: network components belong to practically just one municipality.

\begin{table}[h!]
\caption{Proportion of main municipality within a main connected component.}
\centering
      \begin{tabular}{rrlrr}
        \toprule
         Connected component  & Number of neighborhoods  & Main municipality  & Proportion & Ratio\\ \midrule
        1 & 305 & Monterrey & 220 & 0.72 \\
        2 & 80 & Guadalupe  & 34 & 0.42\\
        3 & 65  & Apodaca & 65 & 1\\ 
        4 & 55 & Guadalupe & 55 & 1\\
        5 & 50 & Santa Catarina  & 48 & 0.96\\
        6 & 23  & Cadereyta  & 23 & 1\\ 
        7 & 19 & Juárez & 19 & 1\\
        8 & 16  & General Escobedo  & 15 & 0.94\\ 
        \bottomrule
      \end{tabular}
      \label{table:connected-components}
\end{table}

This last result is reinforced with Figure~\ref{crime-network}, where each component belongs to one municipality with a high proportion (Table\ref{table:connected-components}). In this sense, a very interesting case is the one corresponding to San Nicolás de los Garza and Guadalupe municipalities, as they share neighborhoods with other violent municipalities in the crime network. On the one hand, Guadalupe has two main connected components associated with it: a  cluster exclusively composed by Guadalupe's neighborhoods, and another with neighborhoods shared with other two municipalities, i.e.\  Monterrey, San Nicolas, Apodaca and Juarez. Finally, there is no main connected component in San Nicolás de los Garza holding a substantial number of neighborhoods.

Probably, both phenomena should be explained by the fact that they are disputed territories or even transit-only places. The second hypothesis may apply better to the case of San Nicolás, as it does not have a main connected component.

The other fact that emerges from the crime network, is that several neighborhoods are in the middle of clearly violent areas but they do not have any casualty during the eight years of data. As an example of this, we point to those neighborhoods between the green, yellow and red clusters in Figure \ref{crime-network}: All these neighborhoods belong to Guadalupe municipality. The gray nodes inside the yellow component, which belong to Monterrey, are also remarkable,  but the most dense set of neighborhoods with no main connected component is the one between the yellow and cyan clusters. This set is part of San Nicolás and despite the fact that (i) this area is surrounded by three crime connected components, and (ii) it also contains non-clustered homicide neighborhoods (black dots), hundreds of neighborhoods are not touched by the homicidal violence.
 
 This last observation coincides with the hypothesis that San Nicolás de los Garza municipality is not a disputed territory, but instead a transit one. This fact is fully inline with the urban environment and its relation with homicides in MMA.
 
 \subsection*{Urban environment and homicidal behavior}
 By observing Figure~\ref{roads}, we may corroborate that the urban environment also shapes the broader distribution of homicides in MMA during the measured period. The central polygon of Monterrey's Downtown encompasses the most violent area of MMA during the whole period (Centro de Monterrey neighborhood). Similarly, the 85th freeway that crosses Monterrey and reaches the southern municipality of Santiago is also a hub in terms of how many homicides occurred on this road.
 
 By observing Figure~\ref{roads} one may notice how many homicidal events occurred on highways. As previously mentioned, it is possible to observe how roads separate places with or without homicides: Apparently the combined political, socio-economical and transportation factors may separate relatively large areas of homicidal violence. 
 
 \section*{Concluding remarks}
 In this work, by means of a network approach, using a unique and carefully curated database of geolocated homicides during a period of eight years, together with map data from Open Street Maps, we have been able to describe the spatiotemporal dynamics of homicidal violence in one of the most important urban areas in Mexico, the Monterrey Metropolitan Area (MMA).
 
 To our knowledge, this is the first time that such an amount of manually curated data is used to construct spatial and temporal networks, to provide insights of how violence increases and decreases as relative to the underlying different social turmoil in Mexico during this time window. 
 
 We have been capable to study at different levels of granularity of temporal (8-year, year, and month), and spatial components (country-level, state-level, municipality, and neighborhood), in order to provide a multi-scale approach that allows to dissect possible explanations behind the violent behavior in urban metropolitan areas, specifically in the case of the war on drugs in Mexico.
 
 Further steps in this regard may focus in performing null models to corroborate whether the geo-socio-political division segregates homicidal regions in MMA. A large effort of several groups should also be made on the data collection and curation.

\section*{Methods}
\subsection*{Data acquisition}
Crime related homicides records for MMA were adquired from EL Norte Mapa del Crimen (2011-2018) (https://gruporeforma.elnorte.com/libre/offlines/mty/mapas/mapadelcrimen2011.htm). Web data extraction was performed for every available year to assemble a single data base: El Norte Data Base (ENDB), with the following variables: date, latitude, longitude, cassualties, title of the newspaper entry reporting the event and associated URL. The total number of observations in the ENDB is 2264. \\
Shapefiles containing neighborhood data for Mexico were downloaded from datamx (http://datamx.io/dataset/colonias-mexico/resource/7b5a3b0a-4405-48d6-a4eb-d9f13bb50d3a). The set was filtered to keep only neighborhoods from the  Nuevo Le\'on. The data set consists of 2691 polygon features with neighborhood names, municipality and geospatial data such as area and location. 
The official number of homicides for Mexico, the state of Nuevo Le\'on and the Monterrey Metropolitan Area (MMA) in the 2000-2018 period was adquired from the National Institute of Statistics and Geography (https://www.inegi.org.mx/sistemas/olap/proyectos/bd/continuas/mortalidad/defuncioneshom.asp?s=est). 
\subsection*{Events mapping}
Entries from the ENDB were mapped into the shapefiles using latitude and longitude coordinates to locate the neighborhood and municipality where each event took place. Entries were kept if they were located inside a neighborhood to filter for urban areas. Mapping was performed using \texttt{pandas} and \texttt{geopandas} packages in Python. After the filter, 2114 entries in the ENDB remained. 
\subsection*{Neighborhood networks}
Adjacent polygons for each neighborhood were determined using \texttt{geopandas}. Each node represents a neighborhood and nodes were linked if both nodes had an event during a determined time window in the ENDB and if they were adjacent polygons in the geospatial data. A network was build using the entire period and yearly networks were also generated. For the spatial evolution of homicides, networks were assembled by counting the number of events that were registered in the same week for two adjacent neighborhoods and by using a one week shifted window. Networks were created using  \texttt{networkx} and visualization and analysis of their structural features were performed using Cytoscape. 
\subsection*{Municipalities correlation}
The entire dataset of causalities aggregated by week was used to calculate the Pearson Correlation Coefficient (PCC) between each pair of municipalities in the MMA. A null-distribution was obtained from a thousand iterations of PCC calculation of the resshufled causalities by week for each municipality. A z-score was assigned to each correlation value by placing the observed PCC in the null-distribution. Correlation heatmap was generated using the \texttt{ComplexHeatmap} package in R. 
\subsection*{Open Street Map}
Data from Open Street Map (OSM) was obtained using the Overpass API for the Monterrey Metropolitan Area (bbox = $(-100.8421,
   25.3217), (-99.5650, 26.0346)$). Nodes and ways were parsed from the data using the \texttt{pyosmium} library, where ways were fined-grained to retrieve only their nodes coordinates. 
   
\section*{List of abbreviations}
\begin{itemize}
\item Monterrey Metropolitan Area: MMA
\item El Norte Data Base: ENDB
\item Enrique Peña Nieto: EPN
\item Felipe Calderón Hinojosa: FCH
\item Open Street Map: OSM
\end{itemize}

\bibliographystyle{unsrt}
\bibliography{references}  


\end{document}